\documentstyle[aps,epsf,twocolumn]{revtex} 
\begin{document}
\draft
\title{\bf Unconditional Pointer States from Conditional Master Equations}
\author{ Diego A. R. Dalvit$^1$ , 
         Jacek Dziarmaga$^{1,2}$, 
         and 
         Wojciech H. Zurek$^1$ }
\address{ 1) Los Alamos National Laboratory, T-6, Theoretical Division, 
MS-B288, Los Alamos, NM 87545 \\
2) Institute of Physics, Jagiellonian University, Krak\'ow, Poland}

\date{June 21, 2000}
\maketitle
\tighten
\begin{abstract} 
{\bf 

When part of the environment responsible for decoherence is used to
extract information about the decohering system, the preferred {\it
pointer states} remain unchanged. This conclusion -- reached for a
specific class of models -- is investigated in a general setting of
conditional master equations using suitable generalizations of
predictability sieve. We also find indications that the einselected
states are easiest to infer from the measurements carried out on the
environment.

}
\end{abstract}

{\bf Introduction.} --- Open quantum systems undergo environment-induced
superselection (einselection) which leads to a preferred
set of quasi-classical pointer states \cite{pointer}. They
entangle least with the environment -- and therefore, lose least information.
Hence, they can be found using predictability sieve, which seeks
states minimizing entropy production \cite{zurek}. 

However, the information
lost to the environment could be, in principle, intercepted and recovered.
Will the preferred states remain at least approximately the same when the
environment is monitored in this fashion?
This is a serious concern, as decoherence is caused by the entanglement
between the system ${\cal S}$ and the environment ${\cal E}$. 
It is well known that a pair of
entangled quantum systems suffers from the basis ambiguity: One can find
out about different states of one of them (e. g., ${\cal S}$) by choosing
a different measurement of the other (e. g., ${\cal E}$ ) 
\cite{pointer}. Thus, basis ambiguity may endanger the definiteness of the
einselected states.

  This issue was pointed out, for example, by Carmichael {\it et al.}
\cite{carm},
who used complete monitoring of the photon environment to develop a
trajectory approach to quantum dynamics \cite{car}. Ref.\cite{carm} 
demonstrated that -- when {\it all} of ${\cal E}$
can be intercepted -- any basis of ${\cal S}$ can be inferred from the
appropriate measurement on ${\cal E}$, so at least in that limit
substantial ambiguity is inevitable. This concern 
is further underscored by the
realization \cite{zurek2} that nearly all of our information comes not
from direct observation of the system, but, rather, by intercepting a
small fraction of (e. g. photon) environment.

  Here we use predictability sieve in combination with the conditional
master equation (CME) \cite{cme} (which obtains when only a part of the
environment -- and not all of it -- is traced out). We show
-- using specific models --
that even when the additional data are taken into account, the
pointer states are unchanged. We demonstrate,
using fidelity, that even when all of ${\cal E}$
is intercepted pointer states are unchanged. Moreover, using specific
models we find indications that -- for an 
observer who acquires the data about the
system indirectly by monitoring the environment -- pointer states are
easiest to discover.

{\bf An example of CME.}--- The master equation for a driven two-level
atom whose emitted radiation is measured by homodyne detection
\cite{car,cme} is an example of CME,

\begin{equation}\label{ME2LA}
d\rho \;=\; d\rho^{\rm{UME}}[\rho] \;+\; d\rho_{\rm{st}}[\rho,N] \;\; ;  
\end{equation}
\begin{eqnarray}
&& d\rho^{\rm{UME}}= 
- i \; dt \; \left[ \Omega\sigma_x , \rho \right] + 
dt \; \left( c \rho c^{\dagger} -
             \frac{1}{2} c^{\dagger}c \rho - 
             \frac{1}{2} \rho c^{\dagger}c  \right),  \nonumber \\
&& d\rho_{\rm{st}}= 
(dN-\overline{dN}) \;
\left( \frac{   (c+\gamma) \rho (c^{\dagger}+\gamma^{\star})  }
            { {\rm Tr}[ (c+\gamma)\rho(c^{\dagger}+\gamma^{\star})]  }
       -\rho         
\right) ,
\label{TWOLEVEL}
\end{eqnarray}
where we set the spontaneous emission time to $1$. We use the It\^{o}
version of stochastic calculus. $\rho$ is a $2\times 2$
density matrix of the atom,
$\Omega$ is a frequency of transitions between the excited and the ground
state driven by a laser beam, $\gamma=R e^{i\phi}$ is the amplitude of
the local oscillator in the homodyne detector and
$c=(\sigma_x-i\sigma_y)/2$ is an annihilation operator. $N_t$ is the
number of photons detected until time $t$. Its increment $dN\in \{0,1\}$
is a dichotomic stochastic process with the average

\begin{eqnarray}\label{dN}
\overline{dN} &=& \eta dt {\rm Tr}
[\rho(c^{\dagger}+\gamma^{\star})(c+\gamma)]
                                                               \nonumber\\
         &=& \eta dt [ R^2 + \langle\sigma_x\rangle R\cos{\phi} 
                           - \langle\sigma_y\rangle R\sin{\phi} 
                           + \langle c^{\dagger}c \rangle        ] \;\;
\end{eqnarray}   
and $\overline{dN^2}=\overline{dN}$. 
The parameter $\eta \in [0,1]$ describes the efficiency of the measurement.
Fully efficient 
measurement ($\eta=1$) occurs when the observer is 
continuously projecting the environment onto a pure state, so that an initial 
pure state of the system remains pure after the measurement
\cite{qoptics}. In the
fully inefficient $\eta=0$ case, the observer has no measurement records or 
ignores them completely. In this case, $d\rho_{\rm{st}}=0$. The equation 
reduces to the unconditional master equation (UME), 
$d\rho=d\rho^{\rm{UME}}$.

The average over realisations
$\overline{d\rho}=d\rho^{\rm{UME}}$, because
$\overline{dN-\overline{dN}}=0$ and $\overline{d\rho_{\rm{st}}}=0$.
$\overline{d\rho}=d\rho^{\rm{UME}}$ is not a special property of
Eqs.(\ref{ME2LA},\ref{TWOLEVEL}) but an axiomatic property of any CME. The
noise average means that we ignore any knowledge about the state of
environment so the state of the system cannot be conditioned by this
knowledge.

  For $R=0$ the measurement scheme is simply a photodetection:
$\overline{dN}$ is proportional to the probability that the atom is in the
excited state. Every click of the photodetector ($dN=1$) brings the atom
to the ground state, from where it is excited again by the laser beam. For
$R \gg 1$ the homodyne photodetector current is a linear function of
$\langle \sigma_x \rangle \equiv {\rm Tr}(\rho\sigma_x)$ for $\phi=0$
($x-$measurement) or of $\langle \sigma_y \rangle$ for $\phi=-\pi/2$
($y-$measurement). These homodyne measurements drive the conditional state
of the atom towards $\sigma_x$ and $\sigma_y$ eigenstates respectively.

{\bf Conditional pointer states are unconditional.} --- 
Are the pointer states of a stochastic CME the same as pointer
states of its corresponding deterministic UME ?  An affirmative
answers requires the
assumption that there is only one kind of environment coupled to
the system and to the detector, but to nothing else, so that the detector
can, in principle, be fully efficient ($\eta=1$) in continuously
projecting the environment onto pure states. This assumption is standard
in quantum optics \cite{qoptics}.

  According to predictability sieve \cite{zurek}, pointer
states minimize the increase of von Neumann
entropy, or, equivalently, the decrease of purity $P={\rm Tr}(\rho^2)$ due to
the interaction with an environment. Suppose that we prepare
a system in a pure state $\rho_0=\rho_0^2$. The noise-averaged initial
rate of purity loss is

\begin{eqnarray}\label{dP}
\overline{dP}_0  \;\equiv\;   
{\rm Tr}(2 \rho_0 \overline{d\rho_0})  \;\;+\;\; 
{\rm Tr}(\overline{d\rho_0 d\rho_0}) \;\;.
\end{eqnarray}

  Any CME can be written in the form of Eq.(\ref{ME2LA}), where $N_t$
would represent a general stochastic process. The stochastic process feeds
the information from measurements of ${\cal E}$ into the conditional
state of ${\cal S}$. The noise-averaged
$\overline{d\rho_0}=d\rho_0^{\rm{UME}}$ depends neither on the
efficiency $\eta$ nor on the kind of measurement we make on ${\cal E}$. 
For a deterministic UME the second term on the RHS of
Eq.(\ref{dP}) would be $O(dt^2)$. For a stochastic CME this second term
gives a contribution proportional to $\eta dt$ which comes from ${\rm
Tr}[\overline{ d\rho_{\rm{st}} d\rho_{\rm{st}} }]$. The manifestly
positive second term reduces the rate of purity loss because a
measurement of ${\cal E}$ tends to purify the conditional state. For
$\eta=1$ the observer gains full knowledge about the environmental state,
the conditional state of the system remains pure all the time, and
$dP_t=0$. Thus we see that for $\eta=1$ the two terms of Eq.(\ref{dP})
should cancel one another. Given that the first and the second term cancel
for $\eta=1$ and that the second term is linear in $\eta$, we can write
the initial purity loss rate as

\begin{equation}\label{dPeta}
\overline{dP}_0 \;=\; 
(1-\eta) \; 
{\rm Tr}(2\rho_0 d\rho_0^{\rm{UME}}) \;\;.
\end{equation}
Up to the prefactor of $(1-\eta)$ this expression is the same as the
corresponding one for the UME. Except for $\eta=1$ we can conclude that
pointer states are the same as those for the UME no matter what the
efficiency is or what kind of measurement is being made.

  When $\eta=1$ we have $\overline{dP}_0=0$ and no preferred pointer
states can be distinguished with the predictability sieve, in accordance
with \cite{carm}. However, even the conditional pure state can drift away
from the free unitary evolution due to the coupling with ${\cal E}$ which 
is measured completely ($\eta=1$). The faster it drifts away the
less predictable the state of the system is. The fidelity with respect to
the initial state is defined as $F_t={\rm Tr}(\rho_0 \rho^{\rm int}_t)$,
where the superscript int refers to interaction picture.
For any $\eta$ the noise-averaged initial decrease of fidelity is
  
\begin{equation}\label{dF}
\overline{dF}_0 \;\equiv\; 
{\rm Tr}(\rho_0 \overline{d\rho_0}) \;=\;
{\rm Tr}(\rho_0 d\rho_0^{\rm{UME}}).
\end{equation}
Thus, UME pointer states maximize fidelity.

Fidelity and purity provide a basis for 
two physically different criteria which lead
to the same unconditional pointer states \cite{diosi}. With the benefit
of the hindsight, these results are not totally unexpected: UME is an average
over CME's, so for linear predictability criteria pointer states should not
change. We have however seen that the same holds for purity, which is non
linear.

  The expressions (\ref{dPeta},\ref{dF}) can be worked out for the example
of the two-level atom master equation. For $\Omega\ll 1$ there is one
pointer state: the ground state. An atom in the ground state cannot change
its state by photoemission and the external driving is slow. In the limit
$\Omega\gg 1$ the externally driven oscillations are much faster than
photoemission. In fact it would be misleading to use Eqs.
(\ref{dPeta},\ref{dF}) as they stand. It is more accurate to 
average them over one period of oscillation:
$\overline{dF}_0 \approx
\frac{\Omega}{2\pi} \int_{0}^{ 2\pi/\Omega  } dt\;
{\rm Tr}[ \rho_0 c^{\dagger}_{\rm{int}} \rho_0 c_{\rm{int}} -
    \rho_0 c^{\dagger}_{\rm{int}} c_{\rm{int}}  ] 
= (-3+x^2_0)/8$.
Here the density matrix is parametrized by
$\rho_t=[I+x_t\sigma_x+y_t\sigma_y+z_t\sigma_z]/2$ with
$x^2_t+y^2_t+z^2_t\leq 1$. Given the last constraint, 
the states with $x=\pm 1$ (eigenstates of the self-Hamiltonian
$\Omega\sigma_x$) are pointer states \cite{wiseman}. 
It should be noted that $\overline{dP}_0$ or
$\overline{dF}_0$ for, say, $y=\pm 1$ states ($\sigma_y$-eigenstates)
is only $50 \%$ worse than for the pointers, for reasons that are
specific to our small system.

{\bf Pointer states are the easiest to find out.} ---
Given the assumptions of the argument above, we have seen that 
pointer states do
not depend on the kind of measurement carried out by the observer on the
environment or on its efficiency. This robustness of pointer states might
convey the wrong impression that all kinds of measurements are equivalent
from the point of view of the observer trying to find out about the system
by monitoring its environment. In what follows we give two examples which
strongly suggest that the measurement of the environment states correlated
with the pointer basis of the system is the most efficient one in gaining
information about the state of the system.

  We begin with the two-level atom. In the limit of $\Omega\gg 1$, the
pointer states are eigenstates of the driving self-Hamiltonian
$\Omega\sigma_x$. For $\eta=0$ the UME has a stationary mixed state
$\rho_{s}=I/2+ O(1/\Omega)$. Suppose that we start monitoring the
environment of the atom at $t=0$ (detectors are turned on at $t=0$, and
$\eta(t)=\eta \theta(t)$). How fast do we find out about ${\cal S}$?  
This can be measured by the purity of the conditional state. For
$\eta\ll 1$ the response of $\rho$ to the switching-on of $\eta$ at
$t=0$ can be described by a small perturbation of the density matrix
$\delta\rho=[\sigma_x\delta x+\sigma_y\delta y+\sigma_z\delta z]/2$ such
that $\rho\approx\rho_s+ \delta\rho$.  The evolution of $\delta\rho$ is
described by

\begin{eqnarray}\label{deltarho}
d\delta x &=& dt(- \delta x/2) + 
dn \left( \frac{2R\cos\phi}{1+2R^2} \right)          
\;,\nonumber\\
d\delta y &=& dt(-\delta y/2-2\Omega \delta z) + 
dn \left( \frac{-2R\sin\phi}{1+2R^2} \right)
\;,\nonumber\\
d\delta z &=& dt(-\delta z+  2\Omega \delta y) + 
dn \left( \frac{-1}{1+2R^2} \right) \;,
\end{eqnarray}
where $dn=dN-\overline{dN}$ and $\overline{dN}=\eta dt (R^2+1/2)$. For
$R\gg 1$ a formal solution of these stochastic differential equations
leads to a noise-averaged purity

\begin{eqnarray}\label{P}
&&\overline{P}(t)  \approx  
{\rm Tr}(\rho_s^2) +  {\rm Tr} (\overline{\delta\rho^2})=
\frac{1}{2} + 
\frac{1}{2} \overline{[ \delta x^2_t + \delta y^2_t + \delta z^2_t ]} =\\
&&  \frac{1}{2}+
    \eta\frac{ R^2\cos^2\phi}{ 1+2R^2  }(1-e^{-t})+ 
    \eta\frac{  1+4R^2\sin^2\phi  }{ 6(1+2R^2) }(1-e^{-3t/2}). \nonumber
\end{eqnarray}
For any time $t>0$ the highest purity is obtained for homodyne ($R\gg 1$)
measurement of $\langle \sigma_x \rangle$ ($\phi=0$). As anticipated, this
is the measurement in the basis of environmental states correlated with
the pointer states of the system. The purity saturates for $t\gg 1$ at

\begin{equation}\label{Pstat}
P_{\infty}=\frac{1}{2}+
          \frac{\eta}{6}\left( 3\cos^2\phi + 2\sin^2\phi \right)\; ,
\end{equation}
for $R \gg 1$.
The small $\eta$ measurements in the pointer state $x-$basis ($\phi=0$)
are only $50\%$ better than in the $y-$basis ($\phi=\pi/2$),
(see Fig.1). As mentioned before, in the two-level atom pointer
states are not well distinguished from the chaff by the predictability
sieve.

\begin{figure}[h]\label{fig1}
\centering \leavevmode
\epsfxsize=7cm
\epsfbox{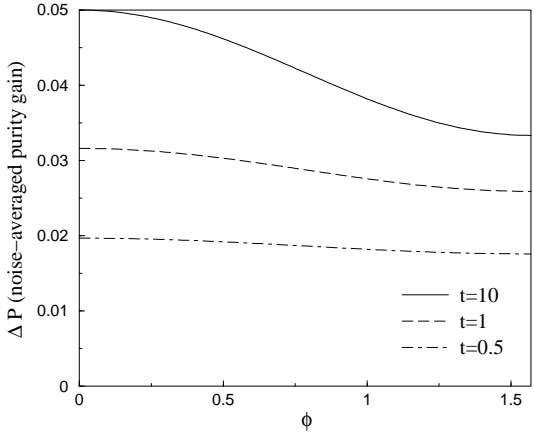}
\caption{ Purity gain, $\Delta P=P-1/2$, as a function of the
homodyne phase $\phi$ according to the formula (\ref{P}) for times: $0.5,
1, 10$.  The parameters were chosen as $\eta=0.1$, $R=100$ (homodyne
limit). }
\end{figure}

  To try with an example known for well distinguished pointer states let
us pick the quantum Brownian motion at zero temperature. We can think of
the environment quanta as phonons. The CME obtains from
Eqs.(\ref{ME2LA},\ref{TWOLEVEL}) by a formal replacement $c\rightarrow a$,
where $a,a^{\dagger}$ are bosonic annihilation/creation operators, 
  
\begin{eqnarray}\label{MEQBM}
d\rho &=& dt \; \left( a \rho a^{\dagger} -
                       \frac{1}{2} a^{\dagger}a \rho - 
                       \frac{1}{2} \rho a^{\dagger}a  \right) 
\nonumber\\
  && + (dN-\overline{dN}) \;
       \left( \frac{   (a+\gamma)\rho (a^{\dagger}+\gamma^{\star})  }
           { {\rm Tr}[ (a+\gamma)\rho(a^{\dagger}+\gamma^{\star}) ] }
                 -\rho         
          \right)  \;\;.
\end{eqnarray}
This equation is valid in the rotating wave approximation. After the
replacement $c\rightarrow a$ in Eq.(\ref{dN}), we get

\begin{equation} 
\overline{dN} = 
\eta dt [ R^2 + \langle a \rangle R e^{-i\phi} +  
          \langle a^{\dagger} \rangle Re^{+i\phi} +  
          \langle a^{\dagger}a \rangle ] \;\;
\end{equation}
For $R=0$ the measurement drives the conditional state to the ground state
of the harmonic oscillator. In the homodyne limit ($R\rightarrow\infty$)
the phonodetector current gives information about the coherent amplitude
$\langle a \rangle$ of the state. The second line of Eq.(\ref{MEQBM})
vanishes for pure coherent states; the conditional state tends to be
localized around coherent states.

For pointer states fidelity loss $\overline{dF}_0={\rm
Tr}[\rho_0 d\rho_0^{\rm{UME}} ]= {\rm Tr} [\rho_0 a^{\dagger} \rho_0 a
-\rho_0 a^{\dagger}a]$ is the least. It vanishes 
if $\rho_0$ is a coherent state,
$\rho_0=| z \rangle\langle z|$ ($a|z\rangle=z|z\rangle$): coherent states
are perfect pointers. In contrast to other states like, say, number
eigenstates initially they do not lose either purity
($\overline{dP}_0=0$) or fidelity ($\overline{dF}_0=0$) (but see
\cite{zhp}).

  We expect homodyne measurement to be provide more 
information than phonodetection. To support this, pick a coherent
state $|z\rangle$. If $r=|z|\gg 1$, then $\langle +z|-z
\rangle\approx 0$ and $a^{\dagger}|z\rangle \approx z^{\star}|z\rangle$.  
In this approximation, a general density matrix in the subspace spanned by
$|\pm z\rangle$ is

\begin{eqnarray}\label{AC}
\rho &=& \frac{1+A}{2} |+z\rangle\langle +z| \; + \;
      \frac{1-A}{2} |-z\rangle\langle -z| \; + \nonumber \\
&&    +C |+z\rangle\langle -z| \; + \;
      C^{\star} |-z\rangle\langle +z| \;.
\end{eqnarray}
Here $A\in[-1,+1]$. Substitution of Eq.(\ref{AC}) into
Eq.(\ref{MEQBM}), and subsequent left and right projections on $|\pm
z\rangle$, give stochastic differential equations for $A$ and $C$. These
equations are most interesting in two limits.  In the
{\it phonodetection limit} ($R=0$) they are
$dA=0$, $dC=-C[2r^2 dt+(dN-\overline{dN})] $.
The off-diagonal $C$ decays after the decoherence time of $1/r^2 \ll 1$.  
$A$ does not change, phonodetection does not produce any purity.
Phonodetection is a very poor choice: by this measurement we learn nothing
about the system!  In the opposite {\it homodyne detection limit}
($R\rightarrow\infty$) the noise $dN-\overline{dN}$ can be replaced (up to
a constant) by a white-noise $dW$ such that $\overline{dW}=0$ and
$\overline{dW^2}=dt$ \cite{ggh}. Again, $C$ decays
after the decoherence time of $1/r^2$. 
Introducing $B=\tanh^{-1}(A)$, defining
a time scale $\tau \equiv 4 t \eta r^2 \cos^2(\phi-\theta)$ 
(here $\theta$ is the phase of the coherent state, $z=r \exp(i \theta)$),
and a noise $d\zeta \equiv  2 \sqrt{\eta} r \cos(\phi-\theta) dW$
($d\zeta^2=d \tau$), we get the following 
Stratonovich stochastic equation

\begin{equation}\label{dB}
\frac{dB}{d\tau} \;=\; \tanh B \;+\; \frac{d \zeta}{d\tau} \;.
\end{equation} 
Suppose
that at $t=0$ we had $A=B=0$ and $C=0$. This is the most mixed state
possible in our subspace. The probability distribution for $A$ at time
$\tau>0$ is

\begin{equation}\label{p}
p(\tau,A) = 
\frac{(2 \pi \tau)^{-1/2}}{(1-A^2)^{3/2}} 
\exp \left( -\frac{\tau}{2}
           -\frac{\ln^2\left(\frac{1+A}{1-A}\right)}{8\tau}
  \right)
\end{equation} 
This distribution is localized at $A=0$ for $\tau=0$ but after a time
scale $\tau \approx 1$ it becomes concentrated at $A=\pm 1$ (see Fig.2).
By these times the conditional state is almost certainly one of the
coherent states $|\pm z\rangle$ and purity is $1$. The asymptotic bimodal
distribution is obtained the fastest for a homodyne tuned to the phase of
the coherent states, $\phi=\theta$. This result is in sharp contrast to
the nil result for phonodetection.  In Fig.3 we plot three realizations of
a stochastic trajectory $A(\tau)$.

\begin{figure}\label{fig2}
\centering \leavevmode

\epsfxsize=7cm
\epsfbox{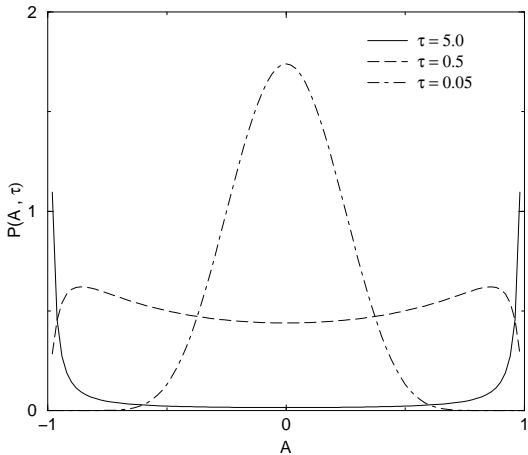}
\caption{ The probability distribution Eq.(\ref{p}) for early
($\tau=0.05$), intermediate ($\tau=0.5$) and late ($\tau=5$)
times.}
\end{figure}

  For any $\eta<1$ purity becomes $1$ after a time proportional to
$1/\eta$.  A patient observer gets full information about the system
monitoring only a small part of the environment:  information about
pointer states is recorded by the environment in a redundant way
\cite{redundant1,redundant2}.

In the above example we assumed that $r=|z|\gg 1$ so that $\langle +z|-z
\rangle\approx 0$ and $a^{\dagger}|z\rangle \approx z^{\star} |z\rangle$.  
This convenient assumption also naturally separates the decoherence and
purification timescales ($\sim 1/r^2$) from the timescale for decay
towards the ground state ($\sim 1$). On the fast timescales $\sim 1/r^2$
we can neglect the decay and that is why our system remains in the $|\pm
z\rangle$-subspace. In this sense our calculation is self-consistent.

{\bf Concluding remarks.} --- The aim of our paper was to study the
issue of the prefered states in the context of conditional master 
equations using the predictability sieve. 
We have shown under reasonable,
but not completely general conditions, that the most
classical states of a system which is being monitored are independent both
of the type of measurement and of the detector efficiency.
Furthermore, we have found indications that the best measurements of the  
environment for gaining information about a system extract data about
its pointer basis.

\begin{figure}\label{fig3}
\centering \leavevmode
\epsfxsize=7cm
\epsfbox{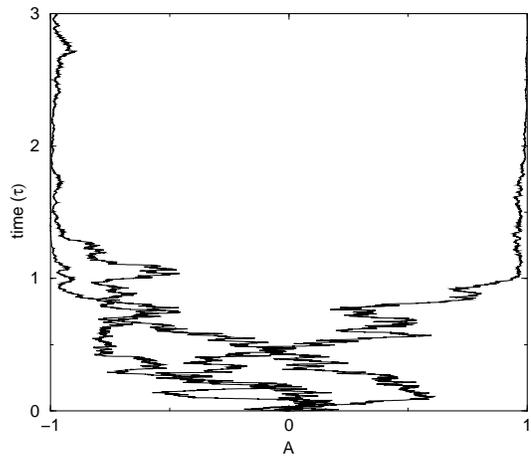}
\caption{Three stochastic trajectories of $A(\tau)$
evolving according to Eq.(\ref{dB}). For any time $\tau$ an average over
such trajectories gives a probability distribution like that in Fig.2.
For late times the trajectories settle down at $A=\pm 1$; the frequency of
jumps between $A=\pm 1$ decays like $\exp(-\tau/2)/\sqrt{\tau}$. }
\end{figure}

{\bf Acknowledgements. } We are grateful to H.M.Wiseman for discussions and
critical comments. This research was supported in part by NSA.

\end{document}